\documentclass[12pt]{article}
\usepackage{epsfig}
\usepackage{a4,isolatin1}
\usepackage{amsmath,amsfonts,latexsym, amssymb}

\newtheorem{satz}{Theorem}[section]
\newtheorem{defi}[satz]{Definition}

\newtheorem{bem}[satz]{Remark}

\newtheorem{koro}[satz]{Corollary}

\newtheorem{ob}[satz]{Observation}

\newtheorem{propo}[satz]{Proposition}

\newcommand{\tit}{\textit}

\newcommand{\N}{\mathbb{N}}

\newcommand{\Z}{\mathbb{Z}}

\begin{document}
\thispagestyle{empty}
\begin{center}
\vspace*{1.0cm}

{\LARGE{\bf Scale Free Small World Networks\\ and the 
Structure of\\ Quantum Space-Time}}

\vskip 1.5cm

{\large {\bf Manfred Requardt}}\\email: requardt@theorie.physik.uni-goettingen.de 

\vskip 0.5 cm 

Institut f\"ur Theoretische Physik \\ 
Universit\"at G\"ottingen \\ 
Tammannstr. 1 \\ 
37077 G\"ottingen \quad Germany

\end{center}

\vspace{1 cm}

\begin{abstract}
  We report on parallel observations in two seemingly unrelated areas
  of dynamical network research. The one is the so-called small world
  phenomenon and/or the observation of scale freeness in certain types
  of large (empirical) networks and their theoretical analysis. The
  other is a discrete cellular network approach to quantum space-time
  physics on the Planck scale we developed in the recent past. In this
  context we formulated a kind of geometric renormalisation group or
  coarse graining process in order to construct some fixed point which
  can be associated to our macroscopic space-time (physics). Such a
  fixed point can however only emerge if the network on the Planck
  scale has very peculiar critical geometric properties which
  strongly resemble the phenomena observed in the above mentioned
  networks. A particularly noteworthy phenomenon is the appearance of
  translocal bridges or short cuts connecting widely separated regions
  of ordinary space-time and which we expect to become relevant in
  various of the notorious quantum riddles.
\end{abstract} \newpage
\setcounter{page}{1}
\section{Introduction}
In this paper we want to report on parallel observations in two, at
first glance, quite unrelated fields of current research. The one is a
discrete network approach to quantum space-time physics, the other is
the analysis of large complicated information networks of interacting
agents, displaying the \tit{small world phenomenon} and/or the
\tit{synchronisation} or \tit{phase locking} among, for example,
coupled non-linear oscillators as they occur in biological and related
model systems. As the literature concerning the latter phenomenon is
huge and as we are planning to discuss the relevance of this
particular phenomenon of synchronisation for our own approach to
quantum gravity elsewhere, we mention only very few but typical papers
(\cite{SS2}, \cite{SM}, \cite{SS3}) and the book \cite{Sync}. 

In the following we will rather concentrate on the parallels between,
on the one hand, the \tit{small world phenomenon} and/or the emergence
of \tit{scale free networks} (\cite{Watts}, \cite{WS}, \cite{Newman}
\cite{NW}, \cite{Barabasi}, \cite{BB}) and, on the other hand, our
dynamical network approach to quantum space-time physics (\cite{Re6},
\cite{Re1}, \cite{Re2}, \cite{Re3}, \cite{Re4}, \cite{Re5},
\cite{Re7}).

The small world phenomenon was, for the first time, observed in
empirical networks while in the past, most of the network modelling
exploited the \tit{random graph concept}. In this latter framework
links are drawn (practically) independently of each other according to
a certain \tit{edge probability}. As a consequence, in general no
particular \tit{near-} or \tit{far-order} does exist. Therefore it
came as quite a surprise that in the real world networks do occur,
which seem to encode a certain (hidden) principle which combines a
sparse wiring (usually implying a large \tit{typical node distance}
and a low \tit{local clustering}) with both a surprisingly high
\tit{local clustering} and a relatively small \tit{average node
  distance}.

On the other hand, (sparse) random graphs, more precisely, random
graphs with a comparable edge probability, have typically both a low
local clustering and, at least in general, a small \tit{diameter} or
average node distance (see section \ref{random}). This observation
suggests that a particular principle is at work, amalgamating these
two seemingly antagonistic properties.

Analysing the underlying laws which may lead to such an interesting
structure, Barabasi et al contributed the concept of \tit{scale free
  networks}, having, for example, a power law \tit{vertex degree
  distribution} (in contrast to random graphs in which the vertex
degree is binomially or Poisson distributed). This notion calls to
mind concepts like critical behavior or self-similarity. While the
small world effect is attributed to the existence of \tit{short cuts}
or \tit{hubs}, scale freeness  is a much stronger property and points
to the existence of a certain \tit{hierarchical organisation} of the
network.

This latter observation establishes the link to our own research in
the field of quantum space-time physics. In the recent past, being
unaware of the possible connections to the small world phenomenon, we
were led to the conclusion that the \tit{geometric renormalisation
  group} or \tit{coarse graining process} we developed within our
discrete network approach, leads only to an interesting large scale
fixed point, we finally want to associate with our macroscopic or
mesoscopic space-time (compared to the primordial Planck scale
regime), if the initial network of relations is in a very particular
\tit{critical state}.

Geometrically this implies that the wiring of the primordial network
is in a certain sense organised in a hierarchical, scale free manner,
which strongly resembles the features described above. It is perhaps
particular noteworthy that in our case this scale freeness is
accomplished by a certain \tit{translocal} character of the wiring
which emerges in the renormalisation process (existence of a hierarchy
of short cuts). This provides also strong clues as to certain
mysteries of the quantum world.

We conclude this introduction with some interdisciplinary
speculation. For biological and many other networks the central
objective is that they function. That is, their organisation should
make them robust and resistent to (at least) local and/or stochastic
failures. If one observes the particular kind of (intricate)
organisation in quite a few real networks we mentioned above, one is
led to surmise that properties like scale freeness or small world
behavior will in fact make them more robust (cf. also \cite{BB}).

If the parallels we invoked regarding a possible similar organisation
of microscopic space-time do not turn out to be unfounded, the same
conclusion can be drawn here. That is, given that microscopic
space-time is a highly complicated dynamical network, wildly
fluctuating on small scales, the organisation principles we are going
to describe in the following may be crucial in preventing such a
structure from becoming chaotic or simply disintegrating into incoherent
pieces. An important role in this context is expected to be played by
the alluded translocal bridges on microscopic scales between
macroscopically widely separated lumps of space-time.
\section{Notions from Graph Theory}
In this section we introduce some terminology and concepts employed in
graph theory and fix the notation. As to the general context see for
example \cite{Bollo2}
\begin{defi}\label{graph1}A simple, countable, labelled, undirected graph, $G$, consists of a countable set of nodes or vertices, $V$, and a set of
  edges, $E$, each connecting two of the nodes. There exist no
  multiple edges (i.e. edges, connecting the same pair of nodes) or
  elementary loops (an edge, starting and ending at the same node). In
  this situation the edges can be described by giving the
  corresponding set of unordered pairs of nodes. The members of $V$
  are denoted by $x_i$, the edges by $e_{ij}$, connecting the nodes
  $x_i$ and $x_j$.
\end{defi}
Remarks: We could also admit a non-countable vertex set. The above
restriction is only made for technical convenience. From a physical
point of view one may argue that the \tit{continuum} or uncountable
sets are idealisations, anyhow. The notions \tit{vertex, node} or
\tit{edge, link} or \tit{bond} are used synonymously. Furthermore, the
labeling of the nodes is only made for technical convenience (to make
some discussions easier) and does not carry a physical meaning. As in
general relativity, all models being invariant under \tit{graph
  isomorphisms} (i.e. relabelling of the nodes and corresponding
bonds) are considered to be physically equivalent.  \vspace{0.3cm}

In the above definition the edges are not directed (but oriented; see
below). In certain cases it is also useful to deal with directed
graphs.
\begin{defi}A directed graph is a graph as above, with $E$ consisting
  now of directed bonds or ordered pairs of nodes. In this case we
  denote the edge, pointing from $x_i$ to $x_j$ by $d_{ij}$. There may
  now also exist the opposite edge, denoted by $d_{ji}$.
\end{defi}
\begin{ob}An undirected graph, as in definition \ref{graph1}, can be
  considered as a particular directed graph with $e_{ij}$
  corresponding to the pair of directed edges, $d_{ij},d_{ji}$.
\end{ob}

In the following we deal, for reasons of simplicity, with
\tit{connected} graphs. We denote the number of vertices and edges by
$n$, $m$, respectively. The maximal possible number of edges over $n$
vertices is $N=n(n-1)/2$. $n$, $m$, are called \tit{order} and
\tit{size} of the graph, $G$. The number of edges, being incident with
a node, $x_i$, is called its \tit{vertex degree}, $k_i$. An \tit{edge
  sequence} or \tit{walk} is a sequence of consecutive edges or nodes,
\begin{equation}(e_{i_1,i_2},e_{i_2,i_3},\ldots,e_{i_k-1,i_k})\;\text{or}\;
  (x_{i_1},\ldots,x_{i_k})     \end{equation}
where the edges or nodes need not be distinct. A \tit{path} is an edge
sequence where no $x_{i_l}$ occurs twice with the possible exception
of $x_{i_1},x_{i_k}$. In the latter case the path is called a
\tit{cycle}.

Between each pair of nodes, $x_i,x_k$, there exists a path, $\gamma$,
of minimal length, $l$, with $l(\gamma)=\#(edges)$, connecting $x_i$ and
$x_k$. This path is called a \tit{geodesic path} and defines a
distance function or metric, $d(x,y)$ on the graph. We have
\begin{equation}d(x_i,x_k)=d(x_k,x_i)>0\;\text{for}\; i\neq k
\end{equation} 
\begin{equation}d(x_i,x_k)\leq d(x_i,x_l)+d(x_l,x_k)     \end{equation}
This metric is called the canonical graph metric. There exist of
course other interesting distance concepts on graphs, cf. for example
\cite{Re1}, where we compared, among other things, the canonical with
the Connes-metric on graphs.

The above metric allows us to introduce a neighborhood concept. We
denote by $\Gamma_l(x_0)$ the set of nodes having exactly distance $l$
from the reference node $x_0$ and by $U_l(x_0)$ the set of nodes with
distance $d(x_0,x_i)\leq l$. The cardinality of $\Gamma_1(x_0)$ is
just the vertex degree of $x_0$. In \cite{Re2} we studied
systematically the scaling behavior of $|\Gamma_l(x)|$ and $|U_l(x)|$
with $l$ (that is, the number of nodes in $\Gamma_l(x),U_l(x)$) and
related it to dimensional concepts of, typically, fractal type. We
argued that it encodes the kind of dimension which turns out to be
relevant in many physical systems.

In \cite{Harary} the sequence of
$d_l(x_0):=|\Gamma_l(x_0)|\;l=1,2,\ldots$ is called the \tit{distance
  degree sequence} relative to node $x_0$, and is denoted by
$dds(x_0)$. Tabulating this for the whole graph, $G$, we get the
\tit{distance distribution}
\begin{equation}dd(G)=(D_1,D_2,\ldots)     \end{equation}
with $D_l$ the number of pairs, $(x_i,x_j)$, having distance equal to
$l$. We have of course
\begin{equation}2D_l=\sum_V d_l(x_i)     \end{equation}
Whereas $dd(G)$ derives from the ensemble of $dds(x_i)$, it is
frequently easier to handle and gives a more compact characteristic of
the global wiring structure of the graph under discussion.

From the above a particularly important graph characteristic can be
derived which is heavily employed in the small world context.
\begin{defi}The mean distance, L(G), of a connected graph is given by
  the average of the distances between the pairs of nodes of G.
\begin{equation}L(G):=N^{-1}\cdot\sum l\cdot D_l\;,\;N=n(n-1)/2  \end{equation}
\end{defi} 

If graphs become very large it is frequently very difficult to
envisage the essentials of the geometric structure of a given graph.
So it is useful to develop more concepts which allow us to encode
typical characteristics of the graph under discussion. A more subtle
concept is the \tit{clique structure} or \tit{clique distribution}.
\begin{defi}[Subsimplices and Cliques]With $G$ a given fixed graph and
  $V_i$ a subset of its vertex set $V$, the corresponding {\em induced
    subgraph} over $V_i$ (that is, its edges being the corresponding
  edges, occurring in $G$) is called a subsimplex or a complete
  subgraph, if all its pairs of nodes are connected by a bond. In this
  class, which is in fact partially ordered, the order being given by
  graph inclusion, there exist certain {\em maximal subsimplices},
  that is, subsimplices so that every addition of another node of the
  underlying graph(together with the respective bonds existing in $G$
  and pointing to other nodes of the selected subset destroys this
  property.  These maximal simplices are usually called {\em cliques}
  in combinatorics (we like to call them also lumps as they are the
  candidates for our construction of {\em physical points}).
  \end{defi}
  
  It has been described in detail in e.g. section 4 of \cite{Re4} how
  these cliques can be constructed in an algorithmic way, starting
  from an arbitrary node. Note in particular that a given node will,
  in general belong to many different (overlapping) cliques or lumps.
  The situation is illustrated by the following picture:
\begin{figure}[h]
\centerline{\epsfig{file=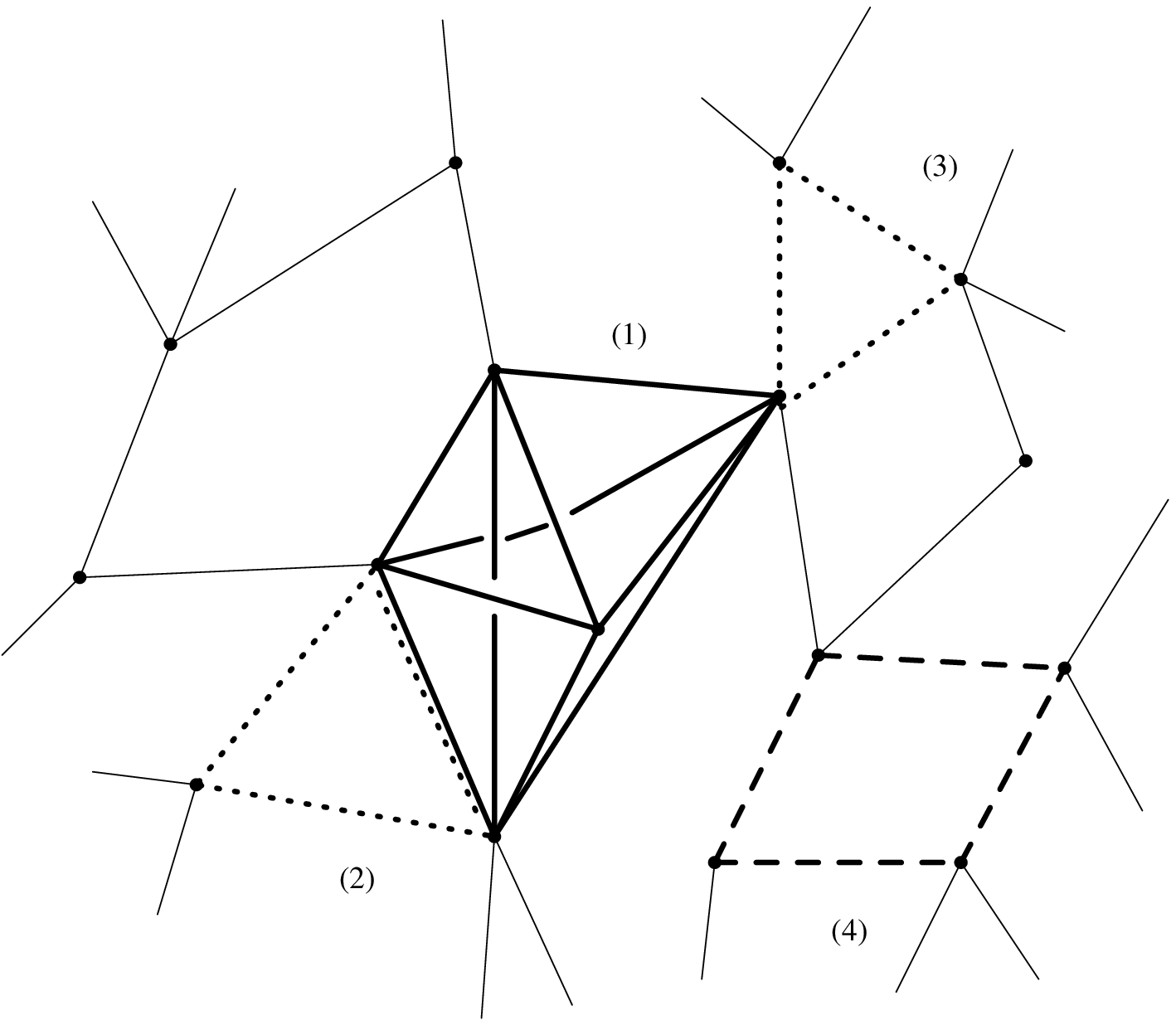,width=6cm,height=6cm,angle=0}}
\caption{}
\end{figure}
In this picture we have drawn a subgraph of a larger graph. $(1)$
denotes a clique, i.e. a maximal subsimplex. Subsets of nodes of such
a clique support subsimplices (called faces in algebraic topology),
the clique being the maximal element in this partial ordered set.
$(2)$ and $(3)$ are other, smaller cliques overlapping with $(1)$ in a
common bond or node. $(4)$ is an example of a subgraph which is not a
clique or subsimplex. Evidently, each node or bond lies in at least
one clique. The smallest possible cliques which can occur in a
connected graph consist of two nodes and the corresponding edge.  The
detailed investigation of the clique structure of graphs was one of
the main topics of \cite{Re3} and earlier in \cite{Re4} and
\cite{Re5}.
\\[0.3cm]
Remark: Note that our definition of a clique (which conforms with
Bollobas') deviates slightly from the one employed by other authors.
Our cliques are the maximal! members in the ascending chains of
complete subgraphs while sometimes the complete subgraphs themselves
are called cliques.  The typical order of cliques turns out to be a
very interesting random variable in random graphs (see below).
\vspace{0.3cm}

Another interesting notion is the \tit{cluster coefficient},
$C_i=C(x_i)$. It is defined by
\begin{equation}C_i:=|E(\Gamma_1(x_i))|/\binom{k_i}{2}
\end{equation}
Here $\binom{k_i}{2}$ is the maximal possible number of edges in
$\Gamma_1(x_i)$ and $|E(\Gamma_1(x_i))|$ the actual
number. Henceforth, $C$ denotes the average over the $C_i$'s.
Further useful concepts are the \tit{average vertex degree} and the
\tit{vertex degree distribution}
\begin{equation}k:=<k>=n^{-1}\cdot\sum k_i=2m/n     \end{equation}
\begin{equation}P(k):=prob(k_i=k)    \end{equation}
where either the probability is taken over a certain ensemble of
graphs or is calculated in a given fixed graph (cf. the end of the
next section).

In this way one can construct a whole bunch of interesting graph
characteristics which are more or less related to each other and their
combination supplying a relatively complete picture of the local and
global structure of a graph. These concepts become particularly
powerful if we combine them with true probabilistic or statistical
concepts. This leads to the definition of a \tit{random graph}.
\section{\label{random}The Concept of a Random Graph}
 One kind of probability space is constructed as follows. Take all
possible labeled graphs over $n$ nodes as probability space $\cal{G}$
(i.e. each graph represents an elementary event). The maximal possible
number of bonds is $N:=\binom{n}{2}$, which corresponds to the unique
{\em simplex graph} (denoted usually by $K_n$). Give each bond the
{\em independent probability} $0\leq p\leq 1$, (more precisely, $p$ is
the probability that there is a bond between the two nodes under
discussion). Let $G_m$ be a graph over the above vertex set, $V$,
having $m$ bonds. Its probability is then
\begin{equation}pr(G_m)=p^m\cdot q^{N-m}\end{equation}
where $q:=1-p$. There exist $\binom{N}{m}$ different labeled
graphs $G_m$, having $m$ bonds, and the above probability is correctly normalized,
i.e.
\begin{equation}prob({\cal G})=\sum_{m=0}^N\binom{N}{m}p^mq^{N-m}=(p+q)^N=1\end{equation}
This probability space is sometimes called the space of {\em
  binomially random graphs} and denoted by ${\cal G}(n,p)$. Note that
the number of edges is binomially distributed, i.e.
\begin{equation}prob(m)=\binom{N}{m}p^mq^{N-m}\end{equation}
and
\begin{equation}\langle m\rangle=\sum m\cdot prob(m)=N\cdot p\end{equation}

The really fundamental observation made already by Erd\"os and R\'enyi (a
rigorous proof of this deep result can e.g. be found in \cite{Bollo2})
is that there are what physicists would call \tit{phase transitions}
in these \tit{random graphs}. To go a little bit more into the details
we have to introduce some more graph concepts.
\begin{defi}[Graph Properties]{\em Graph properties} are certain
  particular {\em random
    variables} (indicator functions of so-called events) on the above
  probability space ${\cal G}$. I.e., a graph property, $Q$, is
  represented by the subset of graphs of the sample space having the
  property under discussion.
\end{defi}
To give some examples: i) connectedness of the graph, ii) existence
and number of certain particular subgraphs (such as subsimplices
etc.), iii) other geometric or topological graph properties etc.

In this context Erd\"os and R\'enyi made the following important
observation.
\begin{ob}[Threshold Function]A large class of {\em graph properties}
  (e.g. the {\em monotone increasing ones}, cf. \cite{Bollo1} or
  \cite{Bollo3}) have a so-called {\em threshold function}, $m^*(n)$,
  with $m^*(n):=N\cdot p^*(n)$, so that for $n\to\infty$ the graphs
  under discussion have {\em property} $Q$ {\em almost shurely} for
  $m(n)>m^*(n)$ and {\em almost shurely not} for $m(n)<m^*(n)$ or vice
  versa (more precisely: for $m(n)/m^*(n)\to \infty\;\text{or}\;0$;
  for the details see the above cited literature). That is, by turning
  on the probability $p$, one can drive the graph one is interested in
  beyond the phase transition threshold belonging to the graph
  property under study. Note that, by definition, threshold functions
  are only unique up to ``factorization'', i.e. $m^*_2(n)=
  O(m^*_1(n))$ is also a threshold function.
\end{ob}

We briefly illustrate the effects of randomisation on some of the
concepts introduced above. We take for example the vertex degree as a
random variable. The probability of a vertex, $x_i$, having vertex
degree $k$ is
\begin{equation}prob(k_i=k)=\binom{n-1}{k}\cdot p^k(1-p)^{n-1-k}\end{equation} 
(the mean value being $p\cdot (n-1)$). In the asymptotic regime of
large $n$, small $k$, $n\cdot p=O(1)$, the vertex degree is
Poisson-distributed (see, for example, \cite{Barabasi}). For a clean
discussion of the assymptotic case of the Poisson distribution see
\cite{Papoulis}. Barabasi et al contrasted this kind of distribution
with so-called \tit{scale free} degree distributions which are
sometimes found in empirical networks (\cite{Barabasi},\cite{BB}).

For the clustering coefficient, defined above, we have in the context
of random graphs the following simple result. As all the edges are
independently distributed with probabiliy $p$ over the random graph,
the distribution is the same for each subclass of nodes, that is, we
have
\begin{equation}C_{rand}=p=k/(n-1)=<m>/N    \end{equation}
In contrast to that typical small world networks may be globally
sparse, i.e. $m/N\ll 1$, but, nevertheless, $C\gg m/N$. That is, in
contrast to randomly wired networks they may display a certain local
order which differs from the order viewed on a more global level.  

Very interesting is the behavior of the average length, $L(G)$, in a
random graph. This random variable, together with the clustering
coefficient, is the pair of graph properties which is primarily
employed to contrast the behavior of random graphs with the wiring
diagrams of so-called small world networks (see \cite{Watts} and the
other literature cited above). 

It is a remarkable (perhaps a little bit counter-intuitive) property
of random graphs that for a large portion of values in the parameter
space, given by the pair $(n,p)$, that is, $p$ not too small, a
typical random graph has diameter less than or equal to two! (for the
details see e.g.  \cite{Bollo1}, some estimates are also given in
\cite{Re4},p.2053f).  To have at all some (weak) scaling with $n$,
$p(n)$ has to vanish for $n\to\infty$ sufficiently strongly without
destroying the connectedness of the graph.

To underpin this qualitative statement, we take over the central
result of chapt. X.2 of \cite{Bollo1} and calculate its asymptotic
behavior for the regime; $n$ large and $p\cdot n\to const.$. We have
for the diameter the approximative result
 \begin{equation}diam\approx const\cdot\log\,n/\log\,pn       \end{equation}
where $p(n-1)=k$ in a random graph, hence
\begin{equation}diam\approx const\cdot \log\,n/\log\,k     \end{equation}
A similar result holds for the average distance, which is, however,
not so easily accessible in general. Strogatz and Watts
(\cite{Watts},\cite{WS}) contrast this weak scaling with the scaling
of, for example, lattice graphs, which are occupying exactly the
opposite end with respect to order or randomness.

The average distance in a $d$-dimensional lattice graph (with, for
simplicity, periodic boundary conditions) can easily be calculated as
follows (cf. also \cite{Re2} for more general scenarios). The
$|\Gamma_l(x_0)|$ of the $dds(x_0)$ of some arbitrary node scale as
$\sim l^{d-1}$. For the average distance relative to $x_0$ we hence
have
\begin{equation}L_{x_0}(G)\sim (n-1)^{-1}\cdot\sum_1^{j_n}l\cdot l^{d-1}    \end{equation}
 with $n\sim j_n^d$. This behaves in leading order as $j_n^{-d}\cdot
 j_n^{d+1}/(d+1)\sim j_n$. That is, $L(G)$ scales linearly with the
 diameter $j_n$.
 
 It is perhaps interesting to compare the two notions, clustering
 coefficient and clique order, introduced above, in a small network
 and a random graph. As cliques are subgraphs of the graphs generated
 by vertices, $x$, and their one-neigborhoods, $\Gamma_1(x)$, it is
 reasonable to calculate the edge probability only with respect to the
 induced subgraphs formed by $\Gamma_1(x)$ and $x$. From the
 clustering coefficient, $C$, and the average vertex degree, $V$, we
 get a corresponding local edge probability:
\begin{equation}p_{loc}=(C\cdot\binom{v}{2}+v)/\binom{v+1}{2}=C-((2C-2)/(v+1))\approx
  C     \end{equation}
for $v$ sufficiently large.

With this $p_{loc}$ we can now calculate the typical clique order,
$r_{loc}$, if we treat the above subgraphs as random graphs. We get
\begin{equation}r_{loc}\approx 2\log(v+1)/\log(p_{loc}^{-1})     \end{equation}
(cf. section 5.1). That is, the clique order scales with $C$ roughly
as $r_{loc}\sim (log\,C^{-1})^{-1}$.

We conclude that, as in a small world network $C$ is larger than
$C_{rand}$, where $C_{rand}$ is the clustering coefficient in a
random graph with the same \tit{global} edge probability, we have
correspondingly $p_{loc}>p_{rand}$ and hence
$r_{loc}>r_{rand}$. Therefore both the clustering coefficient
\tit{and} the clique order are larger in a small world network than in
a true random graph with the same global edge probability.
 
 We close this section with a brief remark about the statistical
 framework, as certain points in this connection are sometimes glossed
 over in the literature. We actually are dealing with two kinds of
 statistics in this enterprise. For one, if we have very large
 networks or graphs, we can apply (practical) statistics within the
 concretely given individual system, that is, perform certain averages
 over nodes, edges and the like. On the other hand, we may prefer to
 study a full ensemble of such graphs, formed according to certain
 statistical or probabilistical principles, an example being the above
 probability space of (binomially distributed) random graphs.
 
 In principle these are different statistical frameworks, but in
 practice they are frequently intermixed. We note in passing that a
 similar philosophy underlies the foundations of statistical
 mechanics. If a system is both sufficiently large and sufficiently
 typical or generic, the differences are expected to be negligible.
 But in any case, it may be wise to remember these (frequently only
 implicit) statistical preassumptions. We made some more detailed
 (physical) remarks about the \tit{statistical hypothesis} in sect.
 3.1 of \cite{Re3}.
\section{Protogeometry and Protodynamics}
We briefly want to motivate why we are modelling the underlying
fabrique of space-time or the quantum vacuum as a relational network
of nodes and links, the geometrical aspects of which can be dealt with
in the context of graphs.

On the one side we
have a working philosophy which is similar to the one, expounded by 't
Hooft in e.g. \cite{Hooft1} to \cite{Hooft3}. That is, we entertain
the idea that for example quantum theory may well emerge as an
\tit{effective (continuum) theory} on the mesoscopic scale of an
underlying discrete more microscopic theory. As we want our underlying
\tit{(pre)geometry} to \tit{coevolve} with the patterns living in this
substratum, we developed the above mentioned generalisation of the
more regular cellular automata.

Another essential property of such \tit{discrete dynamical systems}
is, while the basic ingredients and elementary building blocks are
reasonably simple, their potential for the emergence of very complex
behavior on the more macroscopic scales, thus supporting the
speculation that such systems may be capable of generating viable
continuum theories.

It is now suggestive to regard the edges between pairs of points as
describing their (direct) interaction. This becomes more apparent if
we impose dynamical network laws on these graph structures so that
they become a particular class of discrete dynamical systems.
Henceforth we denote such a dynamical network, which is supposed to
underly our continuous space-time manifold, by $QX$ (``\tit{quantum
  space}''). We want to make the general remark that the \tit{cellular
  networks}, introduced in the following, can either be regarded as
mere models of a perhaps more hypothetical character, encoding, or
rather simulating, some of the expected features of a surmised
\tit{quantum space-time} or, on the other hand, as a faithful
realisation of the primordial substratum, underlying our macroscopic
space-time picture. So far this is a matter of taste.\vspace{0.3cm}

For the time being we choose, to keep matters reasonably simple, a
discrete overall clock-time (not to be confused with the \tit{physical
  time} which is rather supposed to be an emergent and intrinsic
characteristic, related to the evolution of quasi-macroscopic patterns
in such large and intricately wired networks). In principle the
clock-time can also be made into a local dynamical variable.  Cf. also
the complex of investigations grouped around the phenomenon of
synchronization in large populations of coupled oscillators (a small
selection of the existing literature being for example \cite{Watts},
\cite{SS2}, \cite{SM}).  Furthermore, we assume the node set of our
initial network to be fixed and being independent of clock-time (in
contrast to the links). This property may however change if we apply
the renormalisation process which we described in \cite{Re3}. That is,
on the highler levels, the class of \tit{lumps} or \tit{meta-nodes}
may become dependent on time.

On this network we now define a dynamical law or a (clock-) time
evolution.  We assume that each node, $x_i$, or bond, $e_{ik}$,
carries an internal (for simplicity) discrete \tit{state space}, the
internal states being denoted by $s_i$ or $J_{ik}$. In simple examples
we chose for instance:
\begin{equation} s_i \in q\cdot \mathbb{Z}\quad,\quad J_{ik}\in
  \{-1,0,+1\} \end{equation}
with $q$ an elementary quantum of information and
\begin{equation}e_{ki}=-e_{ik}\Rightarrow J_{ki}=-J_{ik}  \end{equation}
In most of the studied cellular automata systems even simpler internal
state spaces are chosen like e.g. $s_i\in \{0,1\}$. This is at the
moment not considered to be a crucial point. The above choice is only
an example.

In our approach the bond states are dynamical degrees of freedom
which, a fortiori, can be switched off or on (see below). Therefore
the \tit{wiring}, that is, the pure \tit{geometry} (of relations) of
the network is a clock-time dependent, dynamical property and is
\tit{not} given in advance.  Consequently, the nodes and bonds are
typically not arranged in a more or less regular array, a regular lattice say,
with a fixed near-/far-order.  This implies that \tit{geometry} will
become to some degree a \tit{relational} (Machian) concept and is no
longer a static background.

As in cellular automata, the node and bond states are updated (for
convenience) in discrete clock-time steps, $t=z\cdot\tau$,
$z\in\mathbb{Z}$ and $\tau$ being an elementary clock-time interval.
This updating is given by some \tit{local} dynamical law (examples are
given below). In this context \tit{local} means that the node/bond
states change at each clock time step according to a prescription with
input the overall state of a certain neighborhood (in some topology)
of the node/bond under discussion.

A simple example of such a local dynamical law we are having in mind
is given in the following definition (first introduced in \cite{Re6}).
\begin{defi}[Example of a Local Law]
At each clock time step a certain {\em quantum} $q$
is exchanged between, say, the nodes $x_i$, $x_k$, connected by the
bond $e_{ik}$ such that 
\begin{equation} s_i(t+\tau)-s_i(t)=q\cdot\sum_k
  J_{ki}(t)\end{equation}
(i.e. if $J_{ki}=+1 $ a quantum $q$ flows from $x_k$ to $x_i$ etc.)\\
The second part of the law describes the {\em back reaction} on the bonds
(and is, typically, more subtle). We assume the
existence of two {\em critical parameters}
$0\leq\lambda_1\leq\lambda_2$ with:
\begin{equation} J_{ik}(t+\tau)=0\quad\mbox{if}\quad
  |s_i(t)-s_k(t)|=:|s_{ik}(t)|>\lambda_2\end{equation}
\begin{equation} J_{ik}(t+\tau)=\pm1\quad\mbox{if}\quad 0<\pm
  s_{ik}(t)<\lambda_1\end{equation}
with the special proviso that
\begin{equation} J_{ik}(t+\tau)=J_{ik}(t)\quad\mbox{if}\quad s_{ik}(t)=0
\end{equation}
On the other side
\begin{equation} J_{ik}(t+\tau)= \left\{\begin{array}{ll} 
\pm1 & \quad J_{ik}(t)\neq 0 \\
0    & \quad J_{ik}(t)=0
\end{array} \right. \quad\mbox{if}\quad
\lambda_1\leq\pm
  s_{ik}(t)\leq\lambda_2 
\end{equation}
In other words, bonds are switched off if local spatial charge
fluctuations are too large or switched on again if they are too
small, their orientation following the sign of local charge
differences, or remain inactive.

Another interesting law arises if one exchanges the role of
$\lambda_1$ and $\lambda_2$ in the above law, that is, bonds are
switched off if the local node fluctuations are too small and are
switched on again if they exceed $\lambda_2$.
\end{defi}
We make the following observation:
\begin{ob}[Gauge Invariance] The above dynamical law depends nowhere on the 
  absolute values of the node ``charges'' but only on their relative
  differences. By the same token, charge is nowhere created or
  destroyed. We have
\begin{equation}\Delta(\sum_{QX}s(x))=0\end{equation}
($\Delta$ denoting the change in total charge of the network between
two consecutive clocktime steps).
To avoid artificial ambiguities we can e.g. choose a fixed reference
level, taking as initial
condition at $t=0$ the following constraint 
\begin{equation}\sum_{QX}s(x)=0\end{equation} 
\end{ob}

We resume what we consider to be the crucial ingredients of network
laws, we are interested in
\begin{enumerate}
\item As in gauge theory or general relativity, our evolution law
  should implement the mutual interaction of two fundamental
  substructures, put a little bit vaguely : ``{\em geometry}'' acting
  on ``{\em matter}'' and vice versa, where in our context ``{\em
    geometry}'' is assumed to correspond in a loose sense to the local
  and/or global array of bond states and ``{\em matter}'' to the
  structure of the node states.
\item By the same token the alluded {\em selfreferential} dynamical
  circuitry of mutual interactions is expected to favor a kind of {\em
    undulating behavior} or {\em selfexcitation} above a return to
  some uninteresting {\em equilibrium state} (being devoid of stable
  structural details), as is frequently the case in systems consisting
  of a single component which directly acts back on itself. This
  propensity for the {\em autonomous} generation of undulation
  patterns is in our view an essential prerequisite for some form of
  ``{\em protoquantum behavior}'' we hope to recover on some coarse
  grained and less primordial level of the network dynamics.
\item In the same sense we expect the large scale pattern of switching-on and
 -off of bonds to generate a kind of ``{\em protogravity}''.
\end{enumerate}
Remark: The above dynamical law shows that bonds with $J_{ik}=0$ at
clock time $t$ do not participate in the dynamics in the next time
step. We hence may consider them as being temporally inactive. The
shape of the network, neglecting all the internal states of the nodes
and bonds together with the inactive bonds we call the \tit{wiring
  diagram}.\vspace{0.3cm}

If one concentrates solely on this \tit{ wiring diagram}, figure 2
(below) describes one clocktime step in the life of a \tit{dynamic
  graph}. In the picture only a small subgraph is shown and the
deletion and creation of edges (that is, elementary interactions among
nodes or possible information channels). The new bonds are represented
as bold lines. It should be emphasized that the graph is \tit{not}
assumed to be a triangulation of some preexisting smooth manifold.
This is emphasized by the existence of edges, connecting nodes which
are not necessarily close with respect to e.g.  the euclidean
distance.
\begin{figure}[h]
\centerline{\epsfig{file=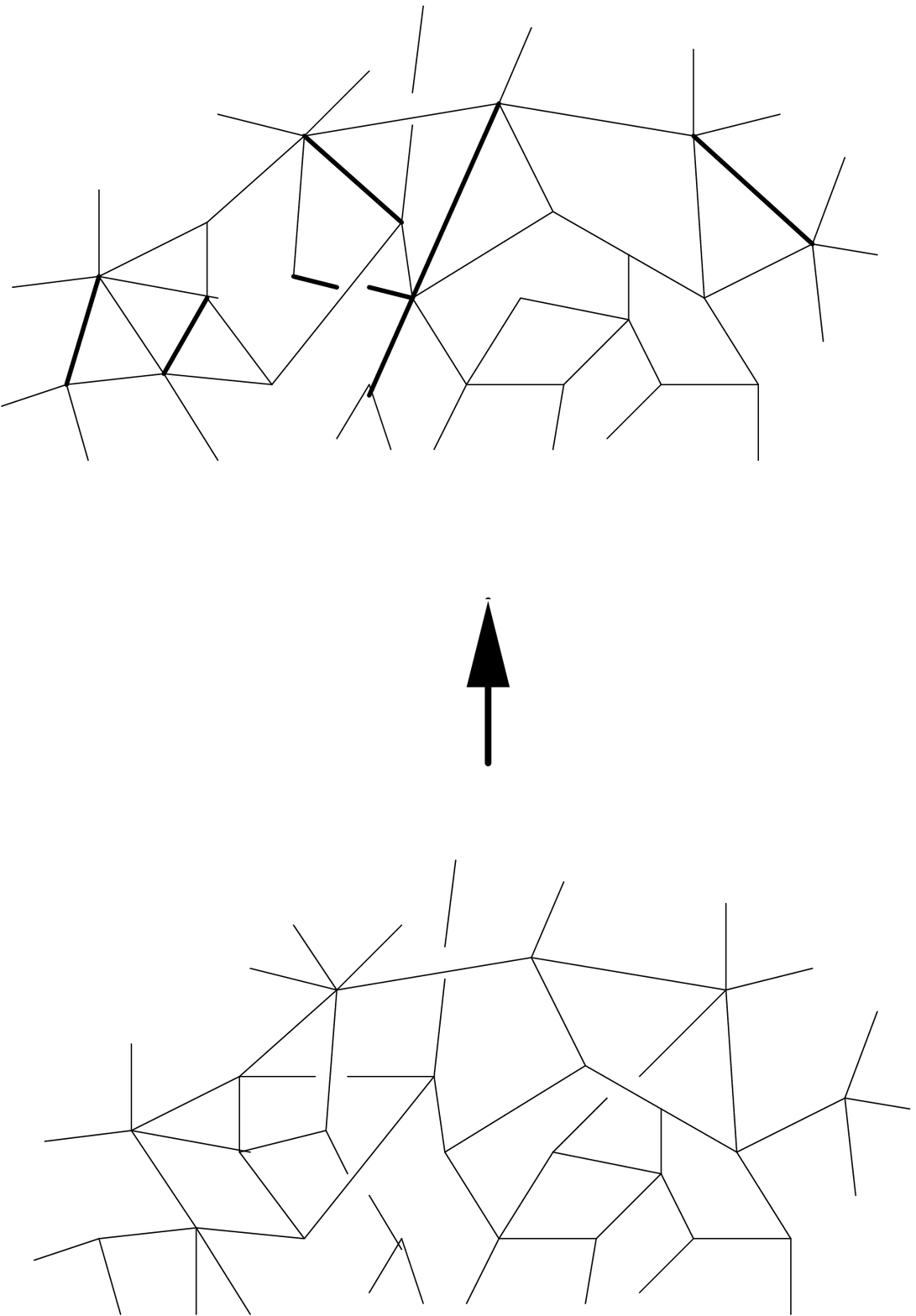,width=7cm,height=7cm,angle=0}}
\caption{}
\end{figure} 
 We recently observed that similar ideas have been entertained within
the framework of cellular automata (see e.g. \cite{Ila1} and
\cite{Ila2}), the models being called \tit{structurally dynamic
  cellular automata} or SDCA. As far as we can see at the moment, the
adopted technical framework is not exactly the same but we think, a
comparison of both approaches should turn out to be profitable.
\begin{bem}We conjecture that such discrete dynamical laws as
  introduced above may be discrete protoforms of the dynamical laws
  governing the arrays of coupled nonlinear oscillators in the papers
  cited previously. This is corroborated by computer simulations on
  arrays of several thousand nodes, performed by us in the past
  (cf. \cite{Nowotny}) which clearly exhibited an amalgamation or
superposition of statistical behavior and a collective undulation
pattern. The second law, described above, has in particular extremely
short transients, reaching a periodic attractor in a very short time,
having the further remarkable property that, given the huge accessible
phase space and the complexity of the network states, it has a period
of only six.  
\end{bem}
\section{The Translocal Depth-Structure of (Quantum) Space-Time}
In this section we want to prepare the stage which will allow us to
relate our own approach to quantum space-time structure with the
small-world network view, being expounded in, on the surface, quite
distinct areas of research. But in order to keep the exposition of the
partly quite intricate technical details within reasonable length, we
will mainly refer, as to the technical details, to the two papers,
\cite{Re3}, \cite{Re7}, and try to give here only the general ideas.

The central picture is that, what we experience as a practically
continuous space-time manifold, will turn out, under sufficient
magnification, as a network of overlapping local clusters or lumps,
being superposed by a, in a measure theoretic sense meager or sparse
second network connecting these local clusters or lumps in a basically
translocal manner. That means, this second network connects local
clusters which may be quite a distance apart with respect to the
metrical structure of the network of underlying lumps. We tried to
make more explicit in \cite{Re3}, \cite{Re4}, \cite{Re7} how this
double structure is expected to go over, on the macroscopic scale of
ordinary space-time physics, into, on the one hand, a smooth local
causal space-time plus, on the other hand, a classically almost hidden
nonlocal network structure which is, due to the weaker and fluctuating
connections, of a more stochastic nature and which we expect to make an
effect in many of the notorious quantum phenomena (note again the
parallels to observations in certain social networks mentioned below).

This overall picture can be made more precise as follows. To construct
this underlying network of lumps, our main tool will be a kind of
\tit{geometric renormalisation group} or systematic \tit{coarse
  graining procedure} which we developed in \cite{Re3}. To put it
briefly, we regard the ordinary space or space-time as a medium having
a rich internal nested fine structure, which is however largely hidden
on the ordinary macroscopic scales due to the usually low level of
resolution of space-time processes as compared to e.g. the Planck
scale. In the process of \tit{coarse graining}, described in the
following, the resolution of the details of space-time is steadily
scaled down from the Planck level to the level of ordinary continuum
physics.

On the deepest level, that is, the Planck scale, proto space-time is
supposed to be a wildly fluctuating network of dynamic relations or
exchange of pieces of information among a given set of nodes. At each
fixed clock-time step there exist certain subclusters of nodes in this
initial network which are particularly densely entangled and the whole
graph can be covered by this uniquely given set of subclusters of
nodes and the respective induced subgraphs (edges given by overlap of
clusters). We dealt with these distinguished clusters of nodes (called
cliques or lumps) in quite some detail in e.g. \cite{Re4} or
\cite{Re5} and define them in the introductory section on graph
concepts. We emphasize the interesting relations to earlier ideas of
Menger, Rosen et al, which have been discussed in \cite{Re5}.

It is fascinating that a similar picture was developed quite some time
ago in mathematical sociology (cf. \cite{Watts} p.14 f and
\cite{Granovetter}). These people developed networks consisting of two
kinds of \tit{ties}, they called (as we did in, for example,
\cite{Re7}) \tit{weak} and \tit{strong} ties. The strong ties define
closely knitted clusters of friends they call \tit{clumps} (similar to
our cliques or lumps) while the weak ties form (nonlocal)
\tit{bridges} to aquaintances who are usually not friends of each
other but are lying in local clumps of their own, with these clumps
non-overlapping with each other. In this work the role of the weak
ties is particularly emphasized, playing a role very similar to the
translocal web in our framework. As to the geometric correspondences
compare this picture with a very similar picture we developed in
section 5.2 (cf. in particular fiure 3), being completely unaware of
the above work.

Technically we need a general principle which allows us to lump
together subsets of nodes, living on a certain level of resolution of
space-time, to construct the building blocks of the next level of
coarse graining (see below). After a series of such coarse graining
steps we will wind up with a nested structure of lumps, containing
smaller lumps and so forth, which, after appropriate \tit{rescaling},
may yield in the end some quasi-continuous but nested structure. This
principle is provided by the following mathematical prescription.
\subsection{The Geometric Renormalization Process}
It is an important observation that in a generic random graph of order
$n$ and edge probability $p$ the order of cliques is concentrated with
very high probability in a relatively small interval $I=(r_0/2,r_0)$;
\begin{equation}r_0\approx 2\log(n)/\log(p^{-1})+ O(\log\log(n))
\end{equation}
see \cite{Bollo1},\cite{Re3},\cite{Re4}. That is, each renormalization
or coarse graining step consists of the following pieces.
\begin{itemize}
\item Starting from a given fixed graph $G=G_0$, defining the level zero,
  pick the (generic) \tit{cliques}, $S_i$, in $G$, their order lying
  in the above mentioned interval, $(r_0/2,r_0)$.
\item These cliques form the new nodes of the \tit{clique-graph},
  $G_{cl}=G_1$ of $G=G_0$. The corresponding new edges are drawn between
  cliques, having a (sufficient degree of) overlap. Size, overlap and
  distribution of cliques in a generic (\tit{random}) graph have been
  analyzed in \cite{Re3},\cite{Re4}. 
\item That is, both \tit{marginal} (i.e. very small) cliques (if they
  do exist at all) and, more importantly (as they are more numerous),
  \tit{marginal} overlaps are deleted. In this respect a
  coarse-graining step includes also a certain \tit{purification} of
  the graph structure.
\end{itemize}

What is considered to be a ``sufficient overlap'' depends of course on
the physical context and the general working philosophy.  A particular
node will in general belong to several, and in the case of densely
entangled graphs to many, cliques. The minimal possible overlap is
given by a single common node. If, on the other hand, the cliques on a
certain level of coarse graining are comparatively large, comprising,
say, typically several hundred nodes, it may be reasonable to neglect
{\em marginal}, i.e. to small, overlaps as physically irrelevant and
define a sufficient degree of overlap to consist of an appreciable
fraction of the typical clique order. The numerical effect of such
choices have been studied in sect.5 of \cite{Re3}.
\begin{defi}We call the graph, defined above, the (purified) clique graph,
  $G_{cl}$, constructed from the initial graph, $G$.
\end{defi}

We note in passing that the robustness of, say, the graph property
\tit{connectedness} under these coarse graining steps has been dealt
with in sect. 5 of \cite{Re3}. We emphasize that our coarse graining
procedure is \tit{universal} in the sense that the \tit{same}
principles are applied on every level of the renormalization
procedure, as the transition from a graph to its clique graph is
always a well defined prescription. In the end, after some rescaling
of the length unit, we hope to arrive at a (quasi-)continuous
manifold, displaying, under appropriate magnification, an intricate
internal fine structure. This should (or rather, can only expected to)
happen if the original network has been in a \tit{critical state} as
will be described in the following.

On each level of coarse-graining, that is, after each renormalisation
step, labelled by $l\in\Z$, we get, as in the block spin approach to
critical phenomena, a new level set of cliques or lumps, $S^l_i$, ($i$
labelling the cliques on renormalisation level $l$), consisting on
their sides of $(l-1)$-cliques which are the $l$-nodes of level $l$,
starting from the level $l=0$ with $G=:G_0$. That is, we have
\begin{equation}S^l_j=\bigcup_{i\in j}
  S_i^{(l-1)}\;,\;S_i^{(l-1)}=\bigcup_{k\in i}
  S_k^{(l-2)}\;\text{etc.}\end{equation}
($i\in j$ denoting the $(l-1)$-cliques, belonging, as meta nodes,
 to the $l$-clique, $S_j$).
These cliques form the meta nodes in the next step.
\begin{defi}The cliques, $S_i^0$, of $G=:G_0$ are called zero-cliques. They
  become the one-nodes, $x_i^1$, of level one, i.e. of $G_1$. The one-cliques,
  $S_i^1$, are the cliques in $G_1$. They become the 2-nodes, $x_i^2$,
  of $G_2$ etc.  Correspondingly, we label the other structural
  elements, for example, 1-edges, 2-edges or the distance functions,
  $d_l(x_i^l,x_j^l)$. These higher-level nodes and edges are also
  called meta-nodes, -edges, respectively.
\end{defi}
\begin{bem}The above construction may  lead to the wrong impression that the
  network becomes sparser after each step. Quite to the contrary, the
  number of cliques in $G_{cl}$ may be much larger than the number of
  nodes in the original graph, $G$ (cf. the table in section 3 of
  \cite{Re3}).  This happens if there is an appreciable overlap among
  the occurring cliques, that is, a given node may belong to many
  different cliques. On the other hand, after several renormalisation
  steps, the picture usually seems to become stable in the generic
  case (see subsect. 5.2 of \cite{Re3}).
\end{bem}

We illustrate the preceding remarks with a couple of numerical
results. Starting, as in \cite{Re3}, from an initial network with the
parameters, $n=10^{100},p=0.7$, which implies $r_0=1291$ and taking as
sufficient overlap of cliques a value of, say, fifty, we get after the
first coarse graining step the new parameters on level
one:\\[0.3cm]
$l=1$: $n_1\approx 10^{10^4}$, $p_1\approx 10^{-7\cdot 10^3}$, $r_1=
3$, average vertex degree $k\approx 10^{0.3\cdot
  10^4}$\\[0.3cm]
We see that after only one step the typical cliques are already very
small. Therefore, in the next step, an overlap greater or equal to one
is appropriate, that is we can use the ordinary clique graph instead
of the purified clique graph and get on the second level:\\[0.3cm]
$l=2$: $n_2\approx 10^{10^4}$ (number of cliques on level one),
$p_2\approx p_1\approx 10^{-7\cdot 10^3}$, $r_2=3$\\[0.3cm]
We convinced ourselves that the qualitative picture will no longer
change under further
renormalisation steps.\\[0.3cm]
Remark: We note that the above choice of parameters is not really
crucial. The qualitative picture will essentially remain the same for
other choices (cf. section 5 of \cite{Re3}).\vspace{0.3cm}

We briefly resume the picture we tried to convey in this
subsection. We argued that, what we regard as the building blocks of
our physical space-time continuum and what we dubbed \tit{physical
  points} in previous work, have actually a nested internal structure
which is built up, starting from the Planck scale, via the
renormalisation steps, described above. 

On the other hand, we want to emphasize that such a continuum as a
\tit{fixed point} or limit state is far from being a quasi automatic
consequence of our procedure. Quite to the contrary, the initial
network has to be in a very peculiar \tit{critical state}, see sect.8
of \cite{Re3}, embodying a kind of seemingly \tit{scale free
  translocal order} as being described, for example, in a different
context by Barabasi et al (\cite{Barabasi}). 
\subsection{The Translocal Network}
This subsection mainly refers to \cite{Re7}, but we deal primarily
with the (random) graph aspects which can be found in sect.6 of
\cite{Re7}. We assume that on a certain scale, $l$, of our
renormalisation process, the network of lumps, $G_l$, is
sufficiently close to the continuous limit manifold, $M$. We take its
meta-nodes, i.e. the cliques, $S_i^{(l-1)}$, of level $(l-1)$ as an
approximation of the physical points of $M$.

This clique graph of level $l$ (i.e. after $l$ coarse graining steps
have been performed) carries a natural graph metric, $d_l(S_i,S_j)$,
which is given by the natural distance between the meta nodes (or
cliques), $S_i,S_j$ (with edges given by sufficient overlap). If one
wants to, one can relate this integer-valued (grainy) distance
functional to a continuous distance function between the corresponding
(fuzzy) physical points, $P_i,P_j$ of the associated continuous
manifold (which, in this approach, may be viewed rather as a mental
construct) with
\begin{equation}d_l(s_i,s_j)\sim d_{man}(P_i,P_j)    \end{equation}
For more details as to these aspects see \cite{Re5}.We want to focus
our attention in the following on another somewhat hidden but very
important aspect of the construction.

By assumption we draw an edge between a pair of cliques, $S_i,S_j$, if
they have a certain degree of overlap of common nodes. This has the
effect that all edges which may exist between nodes lying in
$S_i,S_j$, respectively, are deleted in the next step if these cliques
have empty or too marginal overlap. That is, these cliques, occurring
as meta nodes of the following level are now unrelated in the next
step whereas there may still exist a certain (limited) amount of
information exchange on the preceding, more fine grained levels. In
this sense information will also be coarse grained with only
sufficiently robust information surviving the process.

In subsection 5.2 of \cite{Re3} we made a detailed (numerical)
analysis of these effects of renormalisation. Among other things we
calculated, depending on the edge probability $p$, the typical
cardinality of the \tit{local group} of a given clique, $S_0$, i.e.
the number of cliques having an overlap with $S_0$ bigger than some
prescribed number, the typical cardinality of the cliques on the
consecutive levels, the level-dependent edge probability and vertex
degree and so on. We made the important observation that already after
a few steps the whole picture becomes relatively stationary, implying
that the idea of a stationary limit phase is perhaps not so far-fetched.

What we have discussed so far is the \tit{local structure} or
\tit{near order} aspect of the network which becomes more and more
apparent as a consequence of the consecutive coarse graining steps. We
showed also in \cite{Re3} that this process leads to an unfolding of
the initially densely entangled network towards a network having a
large average distance or diameter similar to a local network in
contrast to a typical random graph.

We infer that in contrast to the initial graph, $G=G_0$, in which a
large portion of the vertices is directly connected, most of the
cliques of $G_0$, i.e. nodes of the first level $G_1$, are no longer
directly connected. On the other hand, many of the nodes lying
\tit{in}, say, the non-overlapping cliques, $S_i,S_j$ are connected by
edges of the initial graph $G_0$.

Another important consequence of our analysis is that practically all
occurring cliques are lying in the above interval $(r_0/2,r_0)$ with
$r_0=1291$ in our example. This implies that essentially every edge in
$G_0$ belongs to at least one such large clique. Put differently, with
$S_i,S_j$ two cliques, having a large distance in $G_1$, and $e_{ij}$
some edge connecting two nodes in $S_i,S_j$ respectively, $e_{ij}$
almost certainly belongs to another large clique of roughly the same
order. This conclusion makes the global picture both intricate and
interesting.

We try to express this situation in the following picture   
\begin{figure}[h]
\centerline{\epsfig{file=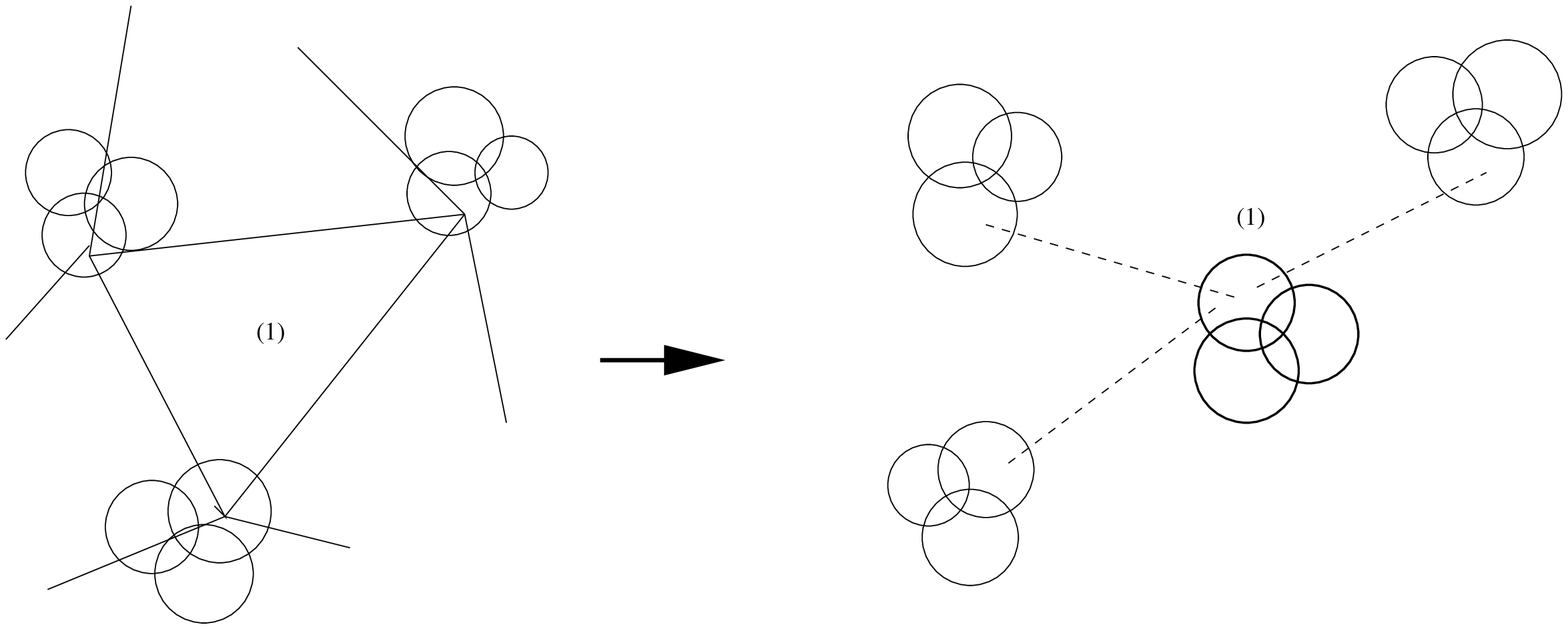,width=12cm,height=6cm,angle=0}}
\caption{}
\end{figure}
The circles denote some generic cliques which are assumed to have
sufficient overlap with (some of) their neighbors. A part of another
clique (denoted by $(1)$), assumed to be of more or less the same
order, but having only \tit{weak} bonds (i.e., weak overlap) with
these possibly widely separated local neighborhood clusters, is
represented by thin lines.

On the other hand, we learned from our numerical estimates (see also
below) that, typically, also clique $(1)$ will have its own local
group, that is, cliques with strong overlap.  In other words, this
particular clique $(1)$ is \tit{almost shurely} the member of another
local group of roughly the same shape, but lying in another region of
the manifold, $M$. This situation is described in the picture on the
right, with the weak bonds between the three local groups, represented
on the left and the clique $(1)$ depicted by dashed lines. The clique
$(1)$ is now represented as a member of another local group of generic
cliques.

Summing up our observations we can conclude that, even if we start
from a densely entangled random graph, we typically arrive after only
a few renormalisation steps at a coarse grained network displaying
both a markedly local behavior in form of strongly coupled local
clusters and a superposed sparser network of a different, translocal
character, with links spreading like a spider web over the whole
underlying local network. We note that these findings exhibit a strong
resemblance to the small world scale free networks discussed
previously. We underpin this speculation with a variety of analytic
results in the following section.
\section{Scale-Free Critical Network States}
As we remarked above, random graphs have a vertex degree distribution
which is of a binomial, and, in a certain limit, of Poisson type. In
contrast to such graphs, Barabasi et al observed that certain classes
of concrete networks are not of this type but instead are of
scale-free type (\cite{Barabasi},\cite{BB}). This means, the vertex
degrees are (asymptotically) distributed according to a \tit{power
  law}
\begin{equation}P(k)\sim k^{-\gamma}   \end{equation} 

Geometrically such a distribution is related to the existence of a
certain portion of \tit{hubs} in the networks, that is, nodes having
an unusually large number of links. However, this is not sufficient in
general. The network has to be hierarchically organised so that these
short cuts do exist on consecutive levels of coarse graining as
described by us in the preceding section. One therefore may suppose
that scale-free networks represent a subclass of small world networks with
clusters in clusters in clusters and so on. In the following we will
corroborate this hypothesis with a number of analytic results.

An important conceptual tool in the analysis of the large scale
behavior of graphs of order, $n$, near or at infinity, is the \tit{distance
degree sequence}, $dds(x)$, relative to an arbitrary but fixed node, $x$
(\cite{Re2},\cite{Re3}). In \cite{Re2} we gave arguments that its
scaling behavior is closely related to a geometric characteristic of
spaces or, rather, systems (in physics) which may be identified with
the concept of \tit{intrinsic dimension}. 
\begin{bem}We use the adjective `intrinsic' to distinguish the concept
  from the more common concept of `embedding dimension'. In contrast
  to the latter it encodes the intrinsic geometrical or relational
  organisation of the system itself and not the lesser important
 structure of the ambient space.
\end{bem}
\begin{defi}[Internal Scaling Dimension] 
  Let $x$ be an arbitrary node of $G$. Let $\#(U_l(x))$ denote
  the number of nodes in $U_l(x)$.We consider the sequence of
  real numbers $D_l(x):= \frac{\ln(\#(U_l(x))}{\ln(l)}$. We say
  $\underline{D}_S(x):= \liminf_{l \rightarrow \infty} D_l(x)$ is the
  {\em lower} and $\overline{D}_S(x):= \limsup_{l \rightarrow \infty}
  D_l(x)$ the {\em upper internal scaling dimension} of G starting
  from $x$. If $\underline{D}_S(x)= \overline{D}_S(x)=: D_S(x)$ we say
  $G$ has internal scaling dimension $D_S(x)$ starting from $x$.
  Finally, if $D_S(x)= D_S$ $\forall x$, we simply say $G$ has {\em
    internal scaling dimension $D_S$}.
\end{defi}
\begin{defi}[Connectivity Dimension] 
  Let $x$ again be an arbitrary node of $G$. Let $\#(\partial
  U_l(x))$ denot the number of nodes in the boundary of $U_l(x)$ (in
  our previous notation $\Gamma_l(x)=\partial U_l(x)$). We
  set $\tilde{D}_l(x) := \frac{\ln(\#(\partial U_l(x))}{\ln(l)} +1$ and
  define $\underline{D}_C(x) := \liminf_{l \rightarrow \infty}
  \tilde{D}_l(x)$ as the {\em lower} and $\overline{D}_C(x) :=
  \limsup_{l \rightarrow \infty} \tilde{D}_l(x)$ as the {\em upper
    connectivity dimension}.  If lower and upper dimension coincide,
  we say $G$ has {\em connectivity dimension} $D_C(x) :=
  \overline{D}_C(x) = \underline{D}_C(x)$ {\em starting from} $x$. If
  $D_C(x) = D_C$ for all $x$ we call $D_C$ simply the {\em
    connectivity dimension} of $G$.
\end{defi}

The two definitions are not strictly equivalent but coincide in the
more regular situations. In the following, for the sake of brevity, we
only use the first notion. Like \tit{fractal dimension} the above
definitions coincide with the usual (embedding) dimension for the more
regular situations like e.g. lattice graphs.

It is remarkable that our concept of graph or network dimension is
stable under a variety of graph transformations or deformations, in
particular local ones (\cite{Re2},\cite{Re3}). In the following we
want to concentrate on one particular aspect, namely the relevance of
a dimensional analysis in connection with \tit{critical network
  states}. These we will associate later with small world scale-free
networks.

In a first step we compare the dimension of a graph, $G$, with its
(unpurified) clique graph, $G_{cl}$, in order to exhibit the
importance of \tit{coarse graining}.
\begin{satz}Assuming that $G$ has dimension $D$ and globally bounded
  node degree, $v_i\leq v<\infty$, we have that $D_{cl}$ also exists
  and it holds
\begin{equation}D_{cl}=D   \end{equation}
Note that this result does hold for the ordinary clique graph, viz.
arbitrary overlap, viz., no purification. In other words, under these
assumptions, the renormalisation steps do not change the graph
dimension.
\end{satz}
The (longer) proof can be found in \cite{Re3}.

This result is reminiscent of a similar observation in statistical
mechanics where the non-coarse-grained Gibbsian entropy happens to be
a \tit{constant of motion}. The same happens here. In the ordinary
clique graph each original bond occurs in at least one clique, i.e.
there is no real (or, more precisely, not enough) coarse graining.

Note that there are two important assumptions underlying the above
result. First, the node degree is assumed to be globally
bounded. Second, there has been no coarse graining. If we allow for
purification, we have the following weaker result.
\begin{koro}For the purified clique graph, with overlaps exceeding a
  certain fixed number, $l_0$, we can only prove
\begin{equation}D_{cl}\leq D\end{equation}
\end{koro}
Having for example the picture in mind, frequently invoked by Wheeler
and others, of a space-time foam, with a concept of dimension
depending on the scale of resolution (see e.g. Box 44.4 on p.1205 in
\cite{Wheeler}), we infer from our above observations that this may
turn out to be both an interesting and not entirely trivial topic. We
have to analyze under what specific conditions the dimension can
actually shrink under coarse-graining, so that we may start from a
very erratic network on, say, the Planck scale, and arrive in the end
at a smooth macroscopic space-time having perhaps an integer dimension
of, preferably, value 4 or so.

While the above corollary seems to allow in principle that space- or
network dimension may become smaller under coarse graining, the
following remarkable result shows that this is not so easily
acchieved. On the other hand, it gives strong clues as to the kind of
(critical) network states which actually do admit a decrease of
dimension under purification.

In \cite{Re2} we investigated the effect of additional edge insertions
in a graph on its dimension. 
\begin{propo}Additional insertions of bonds between arbitrarily many nodes,
  $y,z$, having original graph distance, $d(y,z)\leq k\;,\;k\in\N$
  arbitrary but fixed, do not change $\underline{D}(x)$ or
  $\overline{D}(x)$.
\end{propo}
From this we learn the following. Phase transitions in graphs,
changing the dimension, have to be intrinsically \tit{non-local}. That
is, they necessarily involve nodes, having an arbitrarily large
distance in the original graph. We think, this is a crucial
observation from the physical point of view. It shows that systems
have to be \tit{critical} in a peculiar way, that is, having a lot of
distant correlations or, rather, correlations on all scales (cf. also
Smolins's discussion in e.g. \cite{Smo1} and elsewhere).

In the preceding proposition we made the transition from a graph, $G$,
to a graph $G'$ living on the same node set but having more edges with
the special proviso that edge insertions take only place between pairs
of nodes, $(x,y)$, having
\begin{equation}d_G(x,y)\leq k     \end{equation}
In the purification process we are rather interested in edge
deletions! These two processes are however not! strictly symmetric.

Under edge insertions the distance between nodes does not increase,
i.e.
\begin{equation}G\to G'\Rightarrow d_{G'}(x,y)\leq d_G(x,y)    \end{equation}
On the other hand, edge deletions may lead to
\begin{equation}G'\to G\Rightarrow d_G(x,y)\geq d_{G'}(x,y)    \end{equation}
If we want to employ the above proposition also in the case of edge
deletions, we have to guarantee that edges in $G'$ between nodes,
$x,y$, may be deleted under the proviso that $d_G(x,y)\leq k$. The
condition $d_{G'}(x,y)\leq k$ would not suffice.  
 
If we apply these findings to our renormalisation steps, that is,
passing from a graph to its associated (purified) clique graph, this
implies the following. We saw that assuming a network or graph, $G$,
having a dimension, $D$, the unpurified clique graph still has 
\begin{equation}D_{cl}=D\end{equation}
On the other hand, denoting for the moment the purified clique graph
by $\hat{G}_{cl}$, we have the estimate
\begin{equation}\hat{D}_{cl}\leq D_{cl}=D    \end{equation}

The transition from $G_{cl}$ to $\hat{G}_{cl}$ consists of the
deletion of marginal overlaps among cliques (with the necessary
criteria provided by the physical context). That is, $\hat{G}_{cl}$
lives on the same node set (the set of cliques) but has fewer
(meta)bonds. The above proposition shows that this does \tit{not}
automatically guarantee that we really have 
\begin{equation}\hat{D}_{cl}< D_{cl}   \end{equation}
Quite to the contrary, we learned that this can only be achieved if
the bond deletions happen in a very specific way.

On $G_{cl}$ we have, as on any graph, a natural distance or
neighborhood structure, given by the canonical graph metric,
$d_{cl}(S_i,S_j)$. We thus infer that edge deletions in $G_{cl}$
between cliques which are not very far apart in the final purified
graph $\hat{G}_{cl}$ cannot alter the final dimension of
$\hat{G}_{cl}$. More precisely, only edge deletions between cliques
having distances in $\hat{G}_{cl}$ which approach infinity in a
specific way, can have an effect.

The preceding observation fits into the picture one invokes in the
context of critical behavior and scale freeness in, for example,
statistical mechanics. Furthermore it seems to be closely related to
the kind of scale freeness of networks as observed by Barabasi et al.
In the last section of \cite{Re3} we gave a simple but, as we think,
instructive example in which the effects of edge deletions on all
scales as the origin of dimensional change can explicitly be studied.
The example consists of the one-dimensional (discrete) line, $\Z_1$,
being embedded in $\Z_2$ in a particular way, so that the
corresponding edge deletions lead to a dimensional change from $\Z_2$
to $\Z_1$.

\end{document}